\documentclass[a4paper,11pt]{article}
\pdfoutput=1 
\usepackage{jheppub}
\usepackage{comment}
\usepackage{amsmath,amssymb,amsfonts,amsthm,mathrsfs}
\usepackage{color}

\def\ba{\begin{eqnarray}}
\def\ea{\end{eqnarray}}
\def\bi{\begin{itemize}}
\def\ei{\end{itemize}}

\newcommand{\beq}{\begin{eqnarray}}
\newcommand{\eeq}{\end{eqnarray}}
\newcommand{\be}{\begin{equation}}
\newcommand{\ee}{\end{equation}}

\newcommand{\tr}{\text{Tr}}
\def\D{\mathcal{D}}

\def\bulk{\text{bulk}}
\def\rad{\text{rad}}
\def\A{\mathcal{A}}
\def\F{\mathcal{F}}
\def\I{\mathcal{I}}
\newcommand{\ov}[2]{\overset{\scriptscriptstyle #1}{#2}}
\def\e{\varepsilon}
\def\Lamt{\tilde{\Lambda}}
\def\At{\tilde{\mathcal{A}}}
\def\hard{\textrm{hard}}
\def\soft{\textrm{soft}}
\def\extra{\textrm{extra}}
\def\ext{\textrm{ext}}
\def\w{\omega}
\def\zb{\bar{z}}
\def\L{\mathcal{L}}
\def\Ah{\hat{A}}
\def\J{\mathcal{J}}
\def\Grad{\Gamma^{\textrm{rad}}}
\def\Gext{\Gamma^{\textrm{ext}}}
\def\Oext{\Omega^{\textrm{ext}}}
\def\Orad{\Omega^{\textrm{rad}}}
\def\cov{\text{cov}}
\def\Ggravrad{\Gamma^{\textrm{rad}}_\textrm{grav}}
\def\Ggravext{\Gamma^{\textrm{ext}}_\textrm{grav}}
\def\P{\mathcal{P}}
\def\tf{\textrm{TF}}
\def\S{\mathcal{S}}
\def\tree{\textrm{tree}}
\def\loop{\textrm{1-loop}}

\title{Charge algebra for non-abelian large gauge symmetries at $O(r)$} 
\author[a]{Miguel Campiglia}
\author[a]{Javier Peraza}

\affiliation[a] {\it   Facultad de Ciencias,  Universidad de la Rep\'ublica, \\ Igu\'a 4225, 
              esq. Mataojo, 11400 Montevideo, Uruguay.}
\emailAdd{campi@fisica.edu.uy}
\emailAdd{jperaza@cmat.edu.uy}

\abstract{
Asymptotic symmetries of gauge theories are known to encode infrared properties of  radiative fields. In the context of tree-level Yang-Mills theory, the leading soft behavior of gluons is captured by  large gauge symmetries with parameters that are $O(1)$ in the large $r$ expansion towards null infinity. This relation can be extended to  subleading order provided one allows for  large gauge symmetries with $O(r)$ gauge parameters. The latter, however, violate standard asymptotic field fall-offs and thus their interpretation has remained incomplete. We improve on this situation by presenting a relaxation of the standard asymptotic field behavior that is compatible with  $O(r)$ gauge symmetries at linearized level. We show the extended space admits a symplectic structure on which $O(1)$ and $O(r)$ charges are well defined and such that their Poisson brackets reproduce the corresponding symmetry algebra. 
}

\begin{document} 
\maketitle
\flushbottom

\section{Introduction}

Theories with massless fields  in asymptotically flat spacetimes exhibit a rich structure at null infinity \cite{zrm,aascri,stromlectures} that is reflected in their asymptotic  symmetry groups. Well known examples  are  large gauge symmetries in Yang-Mills (YM) theory \cite{stromYM}, Bondi-Metzner-Sachs (BMS) diffeomorphisms in gravity \cite{bondi,sachs}, and their  higher spin versions  \cite{Campoleoni:2017mbt,Campoleoni:2020ejn}.  

The precise nature of  asymptotic symmetries depends  on the behavior of  fields near null infinity.  Whereas (the Cartesian components of) radiative massless fields typically decay as $1/r$, it is useful in some circumstances  to allow for ``kinematical'' components with slower fall-offs. If achieved consistently, this can lead to an enlargement of the asymptotic symmetry group.\footnote{Relaxing the standard $1/r$ fall-offs is far from straightforward. It typically  leads to inconsistencies, such as divergences in the field's energy flux.  The interplay between consistent boundary conditions and allowed symmetries goes beyond the 4d null infinity case,  see e.g. \cite{Grumiller:2016pqb,Grumiller:2017sjh,Henneaux:2019yax,Compere:2019bua,Aneesh:2021uzk} and references therein.} Interestingly, the resulting enlarged symmetries can  imply non-trivial conservation laws on the original radiative fields. An example of this situation is provided by certain  relaxation on the gravitational field  fall-offs that allows for changes in the leading order ``kinematical'' background metric, yielding  an extension of BMS that includes  ``superrotations''  \cite{BTprl,cl1,dpp}. The resulting conservation law \cite{stromvirasoro} leads to a universal formula for the (tree-level) sub-leading coefficient in the field's frequency expansion \cite{stromcach}.

Inspired by the above gravitational example, it was proposed in \cite{subqed} that large $O(r)$ gauge symmetries can explain similar tree-level gauge theory sub-leading formula  \cite{casali,stromlow}. Whereas the proposal was initially  established in the context of massless scalar electrodynamics, its validity was later extended to more general charged matter \cite{laddhamitra},  higher dimensions, and non-abelian gauge fields \cite{hemitra}.\footnote{In $d=4$ these results are only valid at tree-level, which is the context of the present paper. See \cite{Sahoo:2018lxl} for the sub-leading formula beyond tree level in the abelian case and \cite{alloop,sayali1,sayali2} for possible explanations in terms of asymptotic charges. See  \cite{Bern:2014oka,He:2014bga} for discussion of loop effects in the non-abelian case.}   In these investigations, however, there was no explicit description of the underlying phase space where the symmetries act. In particular, it was not possible to calculate  the algebra of charges. The goal of the present paper is to improve on this situation. For definiteness we will work in the  context of pure YM theory in four spacetime dimensions, although we expect the main ideas should apply  to other settings.

As in the gravitational example, we would like to proceed by first identifying  the appropriate  ``kinematical''  fields that allow for  $O(r)$ gauge symmetries. There is however a major difference between the gravitational and gauge theory  cases: Whereas superrotations form a closed algebra,  $O(r)$ gauge symmetries do not, since their commutator  is  generically $O(r^2)$.  In fact,  once $O(r)$  gauge transformations are allowed, one is forced to include $O(r^n)$ ones for all positive integers $n$. In order to avoid this proliferation, we will  work in an approximation where the   $O(r)$ gauge symmetries are linearized, thus effectively setting to zero the higher order terms.  This restricted setting still allows for interesting structure, in particular regarding the algebra between $O(r^0)$ and $O(r)$ gauge symmetries.   We hope our approximation describes a truncation of an underlying (tree-level) non-linear structure.  The hope is based on  (i) the results of \cite{alsubn,javier} that imply, in the  abelian case, a one-to-one correspondence between $O(r^n)$ large gauge charges and tree-level sub$^{n-1}-$leading formulas \cite{Hamada:2018vrw,Li:2018gnc}, and (ii) the  recently  discovered  \cite{Guevara:2021abz,Strominger:2021lvk} infinite dimensional chiral algebra obeyed by  tree-level (conformally) soft gluons of a given helicity. We  leave for future work the exploration of these would-be higher order non-abelian $O(r^n)$ symmetries.

The organization of the paper is as follows. In the remainder of this section we introduce conventions and notation. In section \ref{radfalloffsec} we discuss the expansion of the gauge field near null infinity under standard radiative fall-offs. In section \ref{reviewchargessec} we review the asymptotic charges associated to the leading and subleading tree-level soft gluon theorems. The core of the paper is  section \ref{secOr}. Here we present the enlarged asymptotic space on which $O(r)$ large gauge transformations act.  We then construct charges associated to $O(r^0)$ and $O(r)$ gauge symmetries on this extended space such that their Poisson brackets reproduce the symmetry algebra. Our approach is similar in spirit to our previous work  \cite{cp} in that we first obtain the charges by requiring certain consistency conditions, and only later define the symplectic structure in a way that is compatible with them. We summarize and highlight open problems in section \ref{finalsec}. In order to facilitate the reading, some of the discussions and  computations are presented in  appendices.

\subsection{Conventions and notation}
We consider pure classical Yang-Mills theory with a matrix group $G$ in 4d flat spacetime.  We denote by $\mathfrak{g}$ the Lie algebra, $[,]$ its Lie bracket, $\A_\mu$ the $\mathfrak{g}$-valued gauge connection and
\be \label{defF}
\F_{\mu \nu} = \partial_\mu \A_\nu - \partial_\nu \A_\mu + [\A_\mu , \A_\nu],
\ee
 the field strength.  The field equations are
\be
\D^\mu \F_{\mu \nu }= \nabla^\mu \F_{\mu \nu}  + [\A^\mu , \F_{\mu \nu}] = 0, \label{fieldeq}
\ee
with $\nabla_\mu$ and $\D_\mu$ denoting the metric and gauge covariant derivatives respectively. Local gauge transformations are parametrized by $\mathfrak{g}$-valued functions $\Lambda$ as
\be
\delta_\Lambda \A_\mu = \D_\mu \Lambda =  \partial_\mu \Lambda + [\A_\mu , \Lambda] . \label{bulkggesym}
\ee
 The ``bulk''  symplectic form is
\be
\Omega^\bulk = -\int_{\Sigma} d S_\mu  \tr (\delta \F^{\mu \nu} \wedge \delta \A_\nu), \label{Obulk}
\ee
where $\tr$ is the matrix trace and the integral is taken over any Cauchy slice $\Sigma$.  The symplectic form can be used  to obtain canonical charges  associated with symmetries. In particular, for the gauge symmetry  \eqref{bulkggesym} one has
\be
 Q^\bulk_\Lambda = - \int_{\Sigma} d S_\mu  \partial_\nu \tr (\Lambda \F^{\mu \nu} ), \label{gralQbulk}
\ee
where  the charge  satisfies $\delta Q^\bulk_\Lambda = \Omega^\bulk(\delta, \delta_\Lambda)$.

To describe the gauge field near future null infinity, we  employ outgoing coordinates $(r,u,x^a)$, where $r$ is the  radial coordinate, $u=t-r$ the retarded time, and $x^a$ coordinates on the celestial sphere. The flat spacetime metric takes the form,
\be
ds^2 = -du^2 -2du dr + r^2 q_{ab}dx^a dx^b,
\ee
where $q_{ab}$ is the  round sphere metric. Points at  null infinity $\I$ will be labeled as $(u,x)$, with  $x = x^a$ denoting a point on the celestial sphere.   The ``bulk'' gauge field $\A_\mu$ induces a gauge field $A_a$ at null infinity,
\be \label{defAa}
A_a(u,x) = \lim_{r \to \infty} \A_a(r,u,x),
\ee
that is unconstrained by the field equations and thus  plays the role of free data.  We will work under the assumption of ``tree-level'' $u \to \pm \infty$ fall-offs
\be
\partial_u A_a(u,x) = O(1/|u|^\infty), \label{falluAa}
\ee
consistent with a  $O(\omega^0)$ subleading behavior in the $\omega \to 0$ frequency expansion of the gauge field.  This  still allows for non-trivial  asymptotic values of $A_a$ at $u = \pm \infty$,
\be
A^\pm_a(x) := \lim_{u \to \pm \infty} A_a(u,x).
\ee

The gauge field near null infinity can be determined in terms of $A_a(u,x)$ by solving the field equations (see e.g. \cite{stromYM,Barnich:2013sxa} and section \ref{radfalloffsec}).  We  denote by $\Gamma^\rad$ the resulting space of gauge fields and write schematically
\be
\Gamma^\rad \approx \{  A_a(u,x) \}. \label{Gammarad}
\ee
Under standard fall-offs, the bulk symplectic form \eqref{Obulk} can be evaluated on the surface $\Sigma \to \I$, leading to the  symplectic form 
\be
\Omega^{\rad} = \int_{\I} \tr (\delta \partial_u A^a \wedge \delta A_a)  du d^2 x  , \label{Orad}
\ee
where $A^a \equiv q^{ab} A_b$ and the determinant $\sqrt{q}$ is implicit in the $d^2 x$ measure. We  refer to the pair  $(\Grad,\Orad)$ as the \emph{radiative phase space}. It is the YM version of the Maxwell and gravity radiative phase spaces introduced in \cite{AS}.

We denote by $D_a$ the gauge-covariant derivative at null infinity,
\be \label{defDa}
D_a := \partial_a + [A_a, ],
\ee
and use $\partial_a$ to denote the sphere-covariant  derivative compatible with $q_{ab}$, i.e. $\partial_c q_{ab}=0$.

\section{YM field near null infinity} \label{radfalloffsec}
In order to specify the $r \to \infty $ expansion of the gauge field we first need to specify a gauge condition. We  will work in  harmonic gauge $\nabla^\mu \A_\mu = 0$, although we expect our results to be valid for  more general gauge choices.  Starting from the standard $O(r^{-1})$ free field fall-offs, one is lead to the following asymptotic expansion:
\be \label{fallrA}
\begin{array}{lll}
\A_r & = &  \frac{1}{r^2}(\ln r \ov{0,\ln}{A}_r+\ov{0}{A}_r ) + \frac{1}{r^3} (\ln r \ov{1,\ln}{A}_r+\ov{1}{A}_r) + o(r^{-3}),  \\
\A_u & = & \frac{\ln r}{r}\, \ov{0,\ln}{A}_u + \frac{1}{r^2}(\ln^2 r \ov{1,\ln^2}{A}_u+\ln r \ov{1,\ln}{A}_u+\ov{1}{A}_u)+ o(r^{-2}), \\
\A_a & = & A_a +  \frac{1}{r}(\ln r \ov{1,\ln}{A}_a  + \ov{1}{A}_a )+o(r^{-1}) ,
\end{array}
\ee
where all coefficients  are functions of $u$ and $x^a$, and  $o(1/r^n)$ denotes quantities decaying faster than $1/r^n$ as $r\to \infty$. We show in appendix \ref{fallrapp}  that \eqref{fallrA} is consistent with  the field equations  and the harmonic gauge condition,
\be \label{botheqs}
\D^\mu \F_{\mu \nu }=0, \quad \nabla^\mu \A_\mu = 0.
\ee
The $r \to \infty$ expansion of  \eqref{botheqs} leads to a hierarchy of equations that can be recursively solved to determine  the coefficients in \eqref{fallrA}  in terms of the free data $A_a$ (modulo integration constants that can be specified by boundary conditions in $u$), see appendix \ref{fallrapp} for details.

The field strength is found to have the following leading $r \to \infty$ behavior,\footnote{The gauge field \eqref{fallrA} appears to  introduce logarithmic terms  that are overleading to those displayed in \eqref{fallrF}. These  however vanish due to the field equations.}
\be \label{fallrF}
\begin{array}{llllll}
\F_{ru} & = &  r^{-2} F_{ru}  + o(r^{-2}),   \quad \quad &  \F_{ra} & = &  r^{-2} F_{ra}  + o(r^{-2}),  \\
& & & & &\\
\F_{ua} & = &  F_{ua} +  o(1),  & \F_{ab} & = & F_{ab} +  o(1).
\end{array}
\ee
From \eqref{defF} and \eqref{fallrA} one has
\be
F_{ua}  = \partial_u A_a , \quad  F_{ab} =  \partial_a A_b - \partial_b A_a + [A_a,A_b].
\ee
$F_{ua}$ plays the role of  asymptotic transverse chromo-electric field   and $F_{ab}$ is the curvature of $A_a$ when viewed as a 2d gauge connection on the celestial sphere. The remaining leading components in \eqref{fallrF} are determined by the asymptotic field equations (see appendix \ref{fallrapp})
\ba
\partial_u F_{ru}  + D^a F_{u a} &= &0 , \label{Frueq}\\
- 2 \partial_u F_{ra} + D_a F_{ru} + D^b F_{ba} &=&0 , \label{Fraeq}
\ea
where we recall that $D_a = \partial_a + [A_a, ]$ and sphere indices are raised with $q^{ab}$.  We shall later integrate these equation using the boundary conditions
\be \label{bdyFcond}
\lim_{u \to + \infty} F_{ru}(u,x)=0, \quad  \lim_{u \to + \infty} F_{ra}(u,x)=0.
\ee
In analogy to the abelian case \cite{alsubn,He:2014cra,eyhe},  we interpret \eqref{bdyFcond}  as due to the absence of massive colored fields. Similar $u \to + \infty$ boundary conditions may also hold for  other coefficients of the field strength, but \eqref{bdyFcond} will suffice for our purposes.

\subsection{Residual large gauge symmetries} \label{resggesymsec}
In order for gauge symmetries \eqref{bulkggesym} to be compatible with the harmonic gauge 
\be
\nabla^\mu \A_\mu = 0, \label{harmgge}
\ee
they must satisfy
\be
 \nabla^\mu \delta_\Lambda \A_\mu = \nabla^\mu \D_\mu \Lambda  =  \square \Lambda +  [\A_\mu , \nabla^\mu \Lambda]  =0. \label{ggeeq}
\ee
This introduces a field-dependence on residual gauge parameters that will be relevant in our later discussion. For the moment, we notice that the commutator of field-dependent gauge transformations can be written as (see e.g. \cite{Barnich:2013sxa}),
\be
[\delta_\Lambda , \delta_{\Lambda'}] \A_\mu = \delta_{[\Lambda, \Lambda']^*} \A_\mu, \label{deltavsmodbrackets}
\ee
where the modified bracket is defined as
\be \label{modbrack}
[\Lambda , \Lambda']^* :=  [\Lambda , \Lambda']+  \delta_{\Lambda} \Lambda' -  \delta_{\Lambda'} \Lambda .
\ee
One can verify that  $[\Lambda , \Lambda']^*$ satisfies  \eqref{ggeeq} provided  $\Lambda$ and $\Lambda'$ do so.

We will be  interested in  large gauge parameters with leading  behavior $O(r^0)$ and $O(r^1)$. We denote these two types of parameters by
\ba
\Lambda^0_\lambda(r,u,x) & \stackrel{r \to \infty}{=}& \lambda(x) + \cdots \label{Lam0}\\
\Lambda^1_\e(r,u,x) & \stackrel{r \to \infty}{=} & r \e (x) + \cdots . \label{Lam1} 
\ea
The  coefficients $\lambda(x)$ and $\e(x)$ are the ``free data'' for the  gauge parameters and the dots represent subleading terms that can be determined   by  solving \eqref{ggeeq}, see appendix \ref{ggeparameterapp}.  Notice that only the $O(r^0)$ gauge parameters are compatible with the radiative fall-offs \eqref{fallrA}. For those one can show that\footnote{It is easy to verify that in this case the leading term of  the  bracket \eqref{modbrack} is given by the ordinary bracket. Since the leading term determines all subleading terms via the gauge parameter equation \eqref{ggeeq}, one concludes both sides of \eqref{Lam0bracket} are equal.}
\be
[\Lambda^0_{\lambda} , \Lambda^0_{\lambda'}]^* = \Lambda^{0}_{[\lambda,\lambda']} \label{Lam0bracket}.
\ee
In  section \ref{secOr} we will present a relaxation of the radiative fall-offs that admit $O(r)$ gauge symmetries to first order in the parameter $\e(x)$. This will allow us to compute the  bracket \eqref{modbrack} between $\Lambda^0_\lambda$ and $\Lambda^1_\e$.

\section{Review of known asymptotic charges} \label{reviewchargessec}
In this section we review the asymptotic charges of YM theory that have been studied in connection with the leading and subleading tree-level soft  factorization formulas. 

Since the seminal work of Strominger \cite{stromYM}  it has been understood that the leading soft gluon factorization can be understood as a conservation law associated to large $O(r^0)$  gauge symmetries (see also \cite{Balachandran:2013wsa,Gonzo:2019fai,Anupam:2019oyi,mitracov,Tanzi:2020fmt}).    On the other hand, the symmetry interpretation  of the subleading factorization  is more subtle: The asymptotic charges are known thanks to the work of  Lysov, Pasterski and Strominger (LPS)  \cite{stromlow},\footnote{The work \cite{stromlow} is in the abelian context but it admits a direct generalization to the non-abelian case; see e.g. \cite{mao,hemitra} and section \ref{LPSsec}.} but it is unclear what  the underlying  symmetry algebra is.  Although progress in this direction has been made from the  perspective of celestial 2d currents \cite{Guevara:2021abz,Adamo:2015fwa,McLoughlin:2016uwa,Banerjee:2020vnt,Pate:2019lpp}, a  ``canonical'' null-infinity picture has been  missing.

\subsection{$O(r^0)$ large gauge charges}
$O(r^0)$ gauge transformations $\delta_{\Lambda^0_\lambda} \A_\mu$  induce an action on the free data $A_a$ that we denote by $\delta^0_\lambda$ and is given by
\be
\delta^0_\lambda A_a =D_a \lambda= \partial_a \lambda + [A_a, \lambda] . \label{del0lam}
\ee
One can verify this action is symplectic wrt $\Omega^\rad$ \eqref{Orad} and satisfies
\be
\Omega^\rad(\delta,\delta^0_\lambda) = \delta Q^{0, \rad}_\lambda
\ee
with
\be
Q^{0, \rad}_\lambda = \int_\I \tr \big( \partial_u A^a D_a \lambda \big) du d^2x. \label{Q0lam}
\ee
An alternative way to obtain this charge is to evaluate  the bulk expression \eqref{gralQbulk} for $\Lambda=\Lambda^0_\lambda$ and $\Sigma \to \I$. Since \eqref{gralQbulk} is a total derivative, this results in a pure boundary term (see e.g. \cite{stromlectures})
\be
Q^{0, \rad}_\lambda = \int_{\I_-}  \tr \big( \lambda(x) F_{ru}(u=-\infty,x) \big) d^2 x, \label{Q0lamspi}
\ee
where $\I_- \approx S^2$ is the $u=-\infty$ boundary of $\I$. The equality between \eqref{Q0lamspi} and \eqref{Q0lam}  follows from the field equation \eqref{Frueq} and the  boundary condition \eqref{bdyFcond}.

\subsection{LPS charges} \label{LPSsec}
The sub-leading soft gluon factorization formula  takes the same form as its abelian counterpart, with color factors replacing abelian charges \cite{casali}. Since we are dealing with pure YM theory, the external colored states are just gluons.  The corresponding creation/annihilation operators are proportional to the negative/positive energy  components of the Fourier transformed asymptotic gauge field,
\be
\Ah_a(\w,x) = \int^\infty_{-\infty} du e^{i \w u} A_a(u,x).
\ee
The non-abelian version of the LPS charges are  parametrized by Lie-algebra valued sphere vector fields $Y^a$ according to
\be
Q_Y =  Q^\soft_Y + Q^\hard_Y
\ee
where
\be
Q^\soft_Y = 2 \lim_{\w \to 0} \partial_{\w} \left( \w \int d^2 x \tr \big(  Y^z \partial^2_z \Ah_{\zb}(\w,x)+ Y^{\zb} \partial^2_{\zb} \Ah_{z}(\w,x)  \big) \right)
\ee
($z$ and $\zb$ are stereographic coordinates on the celestial sphere) and $Q^\hard_Y$ is  defined by\footnote{We use $[ , ]_{\textrm{op}}$ to denote operator commutators in order to distinguish them  from the Lie algebra brackets $[,]$. We have absorbed a factor of $i$ in the definition of $\delta_Y$;  the action of the hard charge is given by $i$ times \eqref{defdelY}.}
\be \label{defQYhard}
[Q^\hard_Y , \Ah_a(\w, x) ]_{\textrm{op}}=  \delta_Y \Ah_a(\w, x),
\ee
\be \label{defdelY}
\delta_Y \Ah_a  := [\partial_a Y^a \partial_\w - \w^{-1} \L_Y, \Ah_a]  =  [\partial_a Y^a , \partial_\w   \Ah_a] - \w^{-1}\big( [Y^b , \partial_b \Ah_a] + [\partial_a Y^b,\Ah_b]\big) .
\ee

Following \cite{stromlow}, one can use these definitions to obtain  expressions of the charges in terms of the radiative data $A_a(u,x)$. One finds
\ba 
 Q^\soft_Y  &= &  -2 \int du d^2 x \, u \tr   Y^z \partial^2_z \partial_u A_{\zb}+ c.c , \label{QYsoft} \\
  Q^\hard_Y   &= &   \int du d^2 x \, u \tr \big( \partial_a Y^a J_u - Y^a \partial_u J_a  \big)  , \label{QYhard}
\ea
where
\be \label{JuJa}
J_u  := [A^a , \partial_u A_a] , \quad   J_z   := 2  q^{z \zb} [ A_z, \partial_z A_{\zb} ].
\ee
The first and second term in \eqref{QYhard} correspond to the first and second term in \eqref{defdelY}.\footnote{The $J_a$ term in \eqref{QYhard}  differs by a total  $u$-derivative from the expression in \cite{stromlow}. Our prescription ensures convergence of the $u$ integral under the assumed fall-offs \eqref{falluAa}.}  As in \cite{stromlow}, the factors $J_u$ and $J_z$ are related to the $O(r^{-2})$ components of the spacetime current, which in our case is just the pure YM ``current'' $\J_\nu =-  \nabla^\mu[\A_\mu,\A_\nu]- [\A^\mu,\F_{\mu \nu}]$.

\section{Extended phase space and $O(r)$ charge algebra} \label{secOr}

In this section we present an extension of the radiative phase space that supports linearized $O(r)$ large gauge symmetries. 

Whereas the standard radiative   space $\Grad$  is parametrized by gauge fields $A_a(u,x)$ at null infinity, the extended space $\Gext$ will include an extra scalar field $\phi(x)$ that can be interpreted as the Goldstone mode associated to $O(r)$ large gauge symmetries (similar to  other known instances of asymptotic symmetries \cite{stromlectures}). 

In section \ref{sec4.1} we present the extended space  and the corresponding action of $O(r^0)$ and $O(r)$ large gauge symmetries, denoted respectively by $\delta^0_\lambda$ and $\delta^1_\e$. Next, we aim to identify the corresponding charges $Q^0_\lambda$ and $Q^1_\e$. Typically this requires knowledge of either Poisson brackets or a symplectic form on $\Gext$. Rather than attempting a first-principles derivation of such structure (which would require a subtle renormalization procedure as in  \cite{superboost,Freidel:2019ohg}), we seek to obtain the charges from a  set of consistency conditions we expect them to satisfy. The conditions are  presented in section \ref{conditionssec}, and the corresponding charges are derived in sections \ref{Q1esec} and \ref{Q0lamsec}. Finally, 
by demanding the charges to arise from a symplectic structure, we  obtain in section \ref{Oextsec} a candidate  symplectic form on $\Gext$. This allows us to realize the $O(r)$ symmetry  algebra obtained in section \ref{sec4.1} as a Poisson bracket charge algebra.

The above ``reverse-logic'' approach of ``charges before symplectic structure'' is inspired by our previous analysis in the gravitational case \cite{cp}, where such strategy was used  to show the existence of a symplectic structure supporting superrotations. In appendix  \ref{gravsec} we discuss in detail the similarities and differences between the YM and gravitational cases.

\subsection{Extended space and $O(r)$ variation algebra} \label{sec4.1}

We would like to minimally relax the radiative fall-offs described in section \ref{radfalloffsec} so as to allow for $O(r)$ gauge transformations. A natural way to proceed is  to apply  all possible $O(r)$ gauge transformations to these radiative fields. Indeed, a similar strategy in the gravitational case leads to an enlargement of the field space that allows for superrotations \cite{Barnich:2010eb,clnewsym,comperevacua,superboost,cp}. As discussed in the introduction, however, in the YM case this procedure cannot be done consistently without allowing for higher order $O(r^n)$ gauge transformations. As a first step, in this paper we   perform a linearized enlargement along the  $O(r)$ gauge direction. We thus consider the  extended  space:
\be
\Gext := \{ \tilde{\A}_\mu = \A_\mu + \D_\mu \Lambda^1_\phi , \quad \A_\mu \in \Gamma^\rad,  \quad \phi \in C^\infty(S^2) \} . \label{defGamma}
\ee
Since $\Gamma^\rad$ is parametrized by fields $A_a(u,x)$,  the extended  space is  parametrized by pairs\footnote{In the analogy with the gravitational case, $\phi$ would correspond to a sphere diffeomorphism labeling the different superrotation sectors, see e.g.  \cite{comperevacua}. Unlike the gravitational case,  we linearize the  finite gauge transformation $\A_\mu \to e^{\Lambda^1_\phi} \A_\mu e^{-\Lambda^1_\phi} + e^{\Lambda^1_\phi} \partial_\mu e^{-\Lambda^1_\phi} \approx  \A_\mu + \D_\mu \Lambda^1_\phi +O(\phi^2)$. See appendix \ref{gravsec} for further details on the gravitational analogue of \eqref{defGamma}.}
\be
 \Gext \approx \{ (A_a(u,x), \phi(x))  \}. \label{Gammavariables}
\ee
By construction, the space \eqref{defGamma} supports the action of  $O(r)$ gauge transformations  \eqref{Lam1}.  In the parametrization \eqref{Gammavariables}, the action is simply given by
\be
\delta^1_\e A_a = 0 , \quad \delta^1_\e \phi = \e. \label{del1e} 
\ee
We emphasize that we are working to first order in  $\phi$ and $\e$. All our expressions should be understood to hold modulo $O(\phi^2)$, $O(\e^2)$ and $O(\phi \e)$ terms.

We next need to specify how $O(r^0)$ gauge transformations  act on $\Gext$.   In the parametrization \eqref{Gammavariables} we define
\be
\delta^0_\lambda A_a = D_a \lambda , \quad \delta^0_\lambda \phi = [\phi,\lambda] , \label{del0lambda}
\ee
leading to an algebra of variations
\be
[\delta^0_{\lambda}, \delta^0_{\lambda'} ] = \delta^0_{[\lambda,\lambda']}, \quad [\delta^1_\e, \delta^0_\lambda] = \delta^{1}_{[\e,\lambda]}, \quad [\delta^1_{\e}, \delta^1_{\e'} ] =0. \label{defalgebra}
\ee
We take \eqref{defalgebra} as the defining relations for the (linearized) $O(r)$ large gauge symmetry algebra.   In appendix \ref{ggeparameterapp} we show how this algebra follows from the bracket  \eqref{modbrack} between $O(r^0)$ and $O(r)$ gauge parameters, and discuss the bulk counterpart of  \eqref{del0lambda}.

\subsection{Conditions on $O(r)$ asymptotic charges} \label{conditionssec}

Our next task is to identify  charges $Q^0_\lambda$ and $Q^1_\e$ on $\Gext$ associated with the symmetries $\delta^0_\lambda$ and $\delta^1_\e$. Since we do not yet know the symplectic structure on $\Gext$, we will find the charges by imposing certain conditions we  expect them to satisfy.  We shall later  determine a symplectic structure on $\Gext$ that is compatible with these conditions.  

Our requirements for the charges are:
\begin{enumerate}
\item  $Q^0_\lambda|_{\Grad} = Q^{0, \rad}_\lambda$ 
\item $Q^1_\e$ is compatible with the tree-level subleading soft gluon factorization
\item  $\delta^0_\lambda Q^1_\e + \delta^1_\e Q^0_\lambda=0$
\item $\delta^0_\lambda Q^1_\e= - Q^1_{[\e,\lambda]}$ 
\end{enumerate}
The first condition requires that when $Q^0_\lambda$ is restricted to  $\Grad \subset \Gext$, one recovers  the standard expression \eqref{Q0lam} for the radiative phase space $O(r^0)$ charge (which is known to encode the leading soft gluon factorization). As we shall discuss, the second condition fixes the   dependence of $Q^1_\e$ on  $A_a(u,x)$ up to (hard) cuadratic order. The third is a necessary condition for the existence of a Poisson bracket realization of the symmetries. The last condition,  probably the least well-motivated one,  requires the charges to reproduce the variation algebra \eqref{defalgebra} without extension terms. 

Our strategy to obtain the charges 
is as follows. It turns out that  conditions 1 and 3 uniquely fix $Q^0_\lambda$ in terms of $Q^{0, \rad}_\lambda$ and $Q^1_\e$, once the latter is known. The most difficult part is then to find   $Q^1_\e$ satisfying conditions 2 and 4. We will thus start by tackling this problem in section \ref{Q1esec}. Once $Q^1_\e$ is known,  we will present the construction of  $Q^0_\lambda$ in section \ref{Q0lamsec}.

\subsection{$Q^1_\e$} \label{Q1esec}

We would like to obtain a  charge $Q^1_\e$ satisfying conditions 2 and 4 above. Condition 2 can be restated as the condition that the Ward identity generated by $Q^1_\e$  is compatible with the one generated by the LPS charge $Q_Y$. In the abelian case,  it was shown in \cite{subqed} that  $Q_Y$ can be understood in terms of an $O(r)$ large gauge charge and its magnetic dual, by  splitting the vector field $Y^a$ into ``electric'' and ``magnetic'' components
\be \label{splitY}
Y_a = \frac{1}{2}(\partial_a \e + \epsilon_{a}^{\ b}\partial_b \mu),
\ee
where $\e(x)$ and $\mu(x)$ are interpreted as the $O(r)$ coefficients of large gauge (and dual gauge) parameters. A first guess could then  be to set $Q^1_{\e}=Q_{Y_a = \partial_a \e/2}$. This however does not satisfy the gauge covariance property required by condition 4. We shall  correct this initial guess so that the resulting charge satisfies 4 without affecting its compatibility with the tree-level soft gluon theorem. We will proceed in two stages: First ``covariantize'' $Q_Y$ and then  consider a gauge covariant version of the splitting \eqref{splitY}.

It is easy to verify that the expression of $Q_Y$ given in Eqs. \eqref{QYsoft}, \eqref{QYhard} is not  gauge covariant at null infinity, i.e.
\be
\delta^0_\lambda Q_Y \neq - Q_{[Y,\lambda]}.
\ee 
Notice however that since $Q_Y$ was read off from a tree-level soft theorem, it only captures terms at most quadratic in $A_a(u,x)$ (see appendix \ref{softthmsapp}). That is,   $Q_Y$ should be understood as giving the $O(A)$ and $O(A^2)$  parts of an asymptotic charge that may contain higher order terms. In addition,  there can also  be   $O(A^2)$ ``soft'' contributions that do not affect the single soft theorem (but which  could leave an imprint in the double-soft behavior). Given this freedom, we now explore the possibility of completing  $Q_Y$  into a gauge-covariant charge. 

A natural way to proceed is to  look for an expression of the charge in terms of the field strength, as in the rewriting of $Q^0_\lambda$ given in Eq. \eqref{Q0lamspi}. Similar  rewritings are known for the abelian subleading charge \cite{stromlow,subqed,alsubn}. A particularly simple expression is one  constructed from  $\F_{ra}$ as we now describe.  The starting point is the asymptotic field equation that relates the $O(r^{-2})$ components of  $\F_{ra}$ and $\F_{ru}$ \eqref{Fraeq},
\be \label{FraFrueq}
- 2 \partial_u F_{ra} + D_a F_{ru} + D^b F_{ba}=0,
\ee
where we recall that 
\be \label{FabitoAa}
F_{ab} = \partial_a A_b - \partial_b A_a + [A_a,A_b],
\ee 
and $F_{ru}$ is determined by Eq. \eqref{Frueq} with boundary condition \eqref{bdyFcond}. Explicitly,
\be \label{FruJu}
F_{ru}   = \partial^a  A^+_a -\partial^a A_a  +   \int^\infty_u J_{u'} du', 
\ee
where $A^+_a(x) = A_a(u=+\infty,x)$ and  $J_u$ is given in \eqref{JuJa}.

From \eqref{FraFrueq} one finds that $F_{ra}=O(u)$ as $u \to - \infty$. The coefficient of the $O(u)$ factor is determined by the $O(1)$ coefficient  of the asymptotic value of the last two terms in \eqref{FraFrueq}. One can then  write an expression for the  finite part of the $u \to - \infty$  asymptotic value of $F_{ra}$, out of which the charge candidate is defined:
\ba 
Q^\cov_Y & : = & \lim_{u \to - \infty} \int d^2 x \tr Y^a \big( 2  F_{ra} - u (D_a F_{ru} + D^b F_{ba}) \big), \\
&= &  \int du d^2 x  u \tr Y^a \partial_u (D_a F_{ru} + D^b F_{ba}) , \label{QYcov}
\ea
where to get the second equality we relied on  the $u \to \infty$ boundary conditions \eqref{bdyFcond} to express the charge as a total $u$-derivative, and  used the field equation \eqref{FraFrueq} to simplify the resulting expression.\footnote{Consistency of \eqref{bdyFcond} with \eqref{FraFrueq} requires that $\lim_{u \to \infty}F_{ab}=0$.}
By construction, the charge expression \eqref{QYcov} is gauge covariant, i.e. it satisfies $\delta^0_\lambda Q^\cov_Y = - Q^\cov_{[Y,\lambda]}$. We now discuss its relation with the LPS charge $Q_Y$. In appendix \ref{covLPSapp} we  show that
\be \label{QYQprimeY}
Q^\cov_Y = Q_Y +  \frac{1}{2} \int du d^2 x u  \tr (\partial_a Y_b- \partial_b Y_a) \partial_u  [A^a,A^b] + \cdots
\ee
where the dots indicate terms that do not affect the tree-level, single-soft gluon behavior. The second term in \eqref{QYQprimeY} is  however incompatible with the subleading soft gluon theorem (see appendix \ref{covLPSapp} for details) and  thus presents an obstruction for the covariantization of $Q_Y$. Fortunately, such term is absent for purely    ``electric'' vector fields $Y_a= \partial_a \e$, which, as described earlier, are the ones relevant for $O(r)$ large gauge charges.\footnote{Eq. \eqref{QYQprimeY} appears to be in conflict with the interpretation of $Q_Y$ as a sum of  electric \emph{and} magnetic $O(r)$ large gauge charges  \cite{subqed}. This  may be related with known  obstructions for a non-abelian extension of  electric-magnetic  abelian duality  \cite{Deser:1976iy}. See \cite{Kapec:2021eug} for a recent discussion of non-abelian magnetic charges at null infinity. \label{magfoot}}

We finally address the non-covariance in  the decomposition \eqref{splitY}. A first guess is to write $Y_a = D_a \e = \partial_a \e +[A_a,\e]$. This however introduces  unwanted quadratic terms in \eqref{QYQprimeY} that would spoil the compatibility of the charge  with the soft theorem. To avoid this problem, we consider a gauge covariant derivative associated to the $u \to - \infty$ asymptotic value of $A_a$,
\be \label{defDminus}
Y_a = D^-_a \e := \partial_a \e + [A_a^-,\e].
\ee
With this definition, the quadratic terms introduced in  \eqref{QYQprimeY} are ``soft'' and hence do not affect the single soft theorem (see appendix \ref{softthmsapp}). We thus define the $O(r)$ large gauge charge as
\ba
Q^1_\e & :=& Q^\cov_{Y_a=D^-_a \e/2} \\
&=&  \int d^2 x  \tr ( \e \pi ) \label{Q1itopi}
\ea
where
\be \label{defpi}
\pi(x) := -\frac{1}{2} \int^\infty_{-\infty} du u \partial_u D^-_a (D^a F_{ru} + D_b F^{ba}) 
\ee
is  a  function of $A_a(u,x)$, due to Eqs. \eqref{FabitoAa}, \eqref{FruJu}, \eqref{defDminus}. Notice that the charge  is independent of the $\phi$ direction  in $\Gext$ \eqref{Gammavariables}. This is because we are working to order $O(\e)=O(\phi)$ and $Q^1_\e$ is already first order in $\e$.

By construction $\pi$ is gauge covariant, in the sense that
\be \label{covariancepi}
\delta^0_\lambda \pi = [\pi,\lambda].
\ee
This immediately implies that  $Q^1_\e$ satisfies the desired covariance property
\be \label{del0lamQ1}
\delta^0_\lambda Q^1_\e = - Q^1_{[\e,\lambda]}.
\ee

We conclude  by emphasizing that our definition of    $Q^1_\e$ does not follow uniquely from requirements 2 and 4 above. 
For instance, one could consider a different prescription for the covariant gradient  in \eqref{defDminus}, or use a different field-strength component as a  starting point (e.g. $\F_{ru}$ instead of $\F_{ra}$, which lead to identical expressions only in the abelian case). All choices would lead to an expression   of the form \eqref{Q1itopi} with slightly different versions of $\pi(x)$. It may be that higher order relations omitted in this work (like  those that would follow from the commutation  between two $O(r)$ charges) could further constraint, and perhaps single out, the form of $\pi(x)$. The discussion in the following sections however is insensitive to the specific form of $\pi(x)$ and only uses the covariance property \eqref{covariancepi}.

\subsection{$Q^0_\lambda$} \label{Q0lamsec}
 We now discuss the extension of $Q^{0, \rad}_\lambda$ to $\Gext$. Given condition 4 is satisfied, condition 3 can be written as
\be \label{condQ0lambda}
  \delta^1_\e Q^0_\lambda = Q^1_{[\e,\lambda]}.
\ee
Since $  \delta^1_\e A_a =0$ and $  \delta^1_\e \phi = \e$, the simplest extension of  $Q^{0, \rad}_\lambda$ that is compatible with  \eqref{condQ0lambda} is
\be \label{defQ0lambda}
Q^0_\lambda = Q^{0, \rad}_\lambda + Q^1_{[\phi,\lambda]}.
\ee
In fact, this is the unique solution to conditions 1 and 3 (for a given $Q^1_\e$). To see why,  consider a different extension $\tilde{Q}^0_\lambda$ and write it as
\be
\tilde{Q}^0_\lambda = Q^0_\lambda + K_\lambda,
\ee
for some function $K_\lambda$ on $\Gext$. Condition 3 then implies
\be
\delta^1_\e K_\lambda =0.
\ee
Given the action of $\delta^1_\e$  \eqref{del1e}, it follows that  $K_\lambda$ must be independent of $\phi$. Thus, $K_\lambda$ must vanish in order to ensure that $\tilde{Q}^0_\lambda|_{\phi=0}=Q^{0, \rad}_\lambda$.

It is interesting to note that due to the gauge covariance of both terms in \eqref{defQ0lambda} it follows that
\be
\delta^0_{\lambda}  Q^0_{\lambda'} = -  Q^0_{[\lambda',\lambda]}.
\ee
Together with  \eqref{del0lamQ1}, this implies  the proposed charges $Q^0_\lambda$ and $Q^1_\e$ reproduce the total variation algebra \eqref{defalgebra}.

\subsection{Extended symplectic form and charge algebra} \label{Oextsec}
We finally present a symplectic form $\Oext$ on $\Gext$ that is compatible with the charges, in the sense that
\be \label{delQsOext}
\delta Q^{0}_\lambda = \Oext(\delta,\delta^0_\lambda) , \quad  \delta Q^{1}_\e = \Oext(\delta,\delta^1_\e) .
\ee
Given the second condition in \eqref{delQsOext} and the form  \eqref{Q1itopi} of $Q^1_\e$ we are lead to define
\be \label{defOext}
\Oext := \Orad + \int d^2 x \tr( \delta \pi \wedge \delta \phi)
\ee
where we recall that
\be
\Omega^{\rad} = \int du d^2 x \tr (\delta \partial_u A^a \wedge \delta A_a)    .
\ee
Indeed, since $\delta^1_\e A_a =0$ (and consequently  $\delta^1_\e \pi =0$)  the only non-trivial contribution to $\Oext(\delta,\delta^1_\e) $ is
\be
\Oext(\delta,\delta^1_\e)  =  \int  d^2 x \tr( \delta \pi \delta^1_\e \phi) = \int  d^2 x \tr( \delta \pi \e ) =\delta  \int  d^2 x \tr( \pi \e )  = \delta Q^1_\e.
\ee
We can now verify that  \eqref{defOext} satisfies the first condition in \eqref{delQsOext}:
\ba
\Oext(\delta,\delta^0_\lambda) & = &  \Orad(\delta,\delta^0_\lambda)  + \int  d^2 x \tr( \delta \pi  \delta^0_\lambda \phi- \delta^0_\lambda \pi \delta \phi) \\
& = & \delta Q^{0, \rad}_\lambda +  \int d^2 x \tr( \delta \pi  [\phi,\lambda]- [ \pi,\lambda ] \delta \phi) \\
& = & \delta Q^{0, \rad}_\lambda + \delta Q^1_{[\phi,\lambda]} = \delta Q^0_\lambda.
\ea
With this symplectic form we can finally realize the relations \eqref{del0lamQ1}, \eqref{condQ0lambda} and \eqref{condQ0lambda} as a Poisson bracket algebra,\footnote{The Poisson bracket between two functions $F$ and $G$ is given by   $\{F,G\} =\Oext(X_G,X_F )$ where $X_F$ is the symmetry transformation generated by $F$, i.e. $\Oext(\delta,X_F ) = \delta F$.}
\be
\{ Q^0_\lambda,Q^0_{\lambda'}\} = Q^0_{[\lambda,\lambda']}, \quad \{ Q^0_\lambda,Q^1_\e\} = Q^1_{[\lambda,\e]} .
\ee

\section{Outlook} \label{finalsec}
The study of asymptotic symmetries in gauge and gravitational theories has proven to be a useful source of information on their infrared properties. The nature of these symmetries, however, depends crucially on the subtle problem of boundary conditions imposed on the fields. In this work, we have proposed an enlargement of the radiative phase space of classical Yang-Mills theory so that it can support linearized $O(r)$ gauge symmetries. The extended space is parametrized by the standard asymptotic gauge field plus a ``Goldstone mode''  that transforms inhomogenously under linearized $O(r)$ symmetries. We showed this extended space admits a symplectic structure such that the charges associated to the $O(r^0)$ and $O(r)$ gauge symmetries are compatible with known tree-level soft gluon theorems, and such that their Poisson brackets reproduce the variation algebra. 

There are several future directions that appear worth pursuing. 

On the one hand, we hope our results are the first order approximation of a higher order symmetry algebra, at least within the tree-level theory. The next order would include linearized $O(r^2)$ gauge transformations and $O(r)$ ones at second order. The former would be related to the (partial) sub-subleading soft gluon factorization \cite{Hamada:2018vrw}. In this context, it would be important to make contact with the ``celestial'' 2d CFT approach to symmetries \cite{Cheung:2016iub,Nande:2017dba,He:2015zea,Himwich:2019dug,Fan:2019emx,Kalyanapuram:2021tnl,Raclariu:2021zjz,Pasterski:2021rjz}, which naturally incorporates higher order factorization formulas satisfying a rich algebraic structure \cite{Guevara:2021abz,Jiang:2021ovh}.

On the other hand, staying within the algebra at $O(r)$, it would be exciting if one could  extend the analysis to include loop corrections to the soft gluon factorization formulas, see e.g. \cite{Bern:2014oka}. It is interesting to note that the algebra consistency  lead us to include higher order terms in the charges that would be sensitive to such loop effects. 
The problem however would be quite challenging, in particular due the appearance of infrared divergences. Here again it may prove useful to establish contact with the celestial CFT methods, which appear well suited for the incorporation of these effects \cite{Albayrak:2020saa,Gonzalez:2020tpi,Gonzalez:2021dxw,Magnea:2021fvy}.

Much of the inspiration for the present work stems from the study of asymptotic symmetries in gravity. In this regard, it has been proposed \cite{subsub} that asymptotic diffeomorphisms  generated by certain $O(r)$ sphere-vector fields are behind the sub-subleading soft graviton factorization \cite{stromcach}. Perhaps a similar strategy as the one presented here could be used to identify an extended space supporting such singular transformations.

\section*{Aknowledgements}
We would like to thank Alok Laddha for illuminating discussions and for his feedback on the manuscript. We acknowledge  support from PEDECIBA and from ANII grant FCE-1-2019-1-155865. 

\appendix

\section{Asymptotic gauge field} \label{fallrapp}
In this appendix we  show that  asymptotic behavior
\be \label{fallrAapp}
\begin{array}{lll}
\A_r & = &  \frac{1}{r^2}(\ln r \ov{0,\ln}{A}_r+\ov{0}{A}_r ) + \frac{1}{r^3} (\ln r \ov{1,\ln}{A}_r+\ov{1}{A}_r) + o(r^{-3}),  \\
\A_u & = & \frac{\ln r}{r}\, \ov{0,\ln}{A}_u + \frac{1}{r^2}(\ln^2 r \ov{1,\ln^2}{A}_u+\ln r \ov{1,\ln}{A}_u+\ov{1}{A}_u)+ o(r^{-2}), \\
\A_a & = & A_a +  \frac{1}{r}(\ln r \ov{1,\ln}{A}_a  + \ov{1}{A}_a )+o(r^{-1}) ,
\end{array}
\ee
is consistent with the field and gauge fixing equations \eqref{botheqs}. As it will become clear,  the appearance of logarithmic terms is forced upon us due to the interaction terms.  In the expansion above we have  used a residual $O(r^{-1})$  gauge freedom to set to zero the $O(r^{-1})$ component of $\A_u$.

The above gauge field expansion leads to the following field strength expansion,
\be \label{fallrFapp}
\begin{array}{lll}
\F_{ru} & = &  \frac{1}{r^2}(\ln r \ov{0,\ln}{F}_{ru}+\ov{0}{F}_{ru} ) + \frac{1}{r^3}(\ln^2 r \ov{1,\ln^2}{F}_{ru}+\ln r \ov{1,\ln}{F}_{ru}+\ov{1}{F}_{ru}) + o(r^{-3}),  \\
\F_{ra} & = &  \frac{1}{r^2}(\ln r \ov{0,\ln}{F}_{ra}+\ov{0}{F}_{ra} ) + o(r^{-2}),  \\
\F_{ua} & = &  \ov{0}{F}_{ua} + \frac{1}{r}(\ln r \ov{1,\ln}{F}_{ua}+\ov{1}{F}_{ua} ) + o(r^{-1}),  \\
\F_{ab} & = & F_{ab} +  \frac{1}{r}(\ln r \ov{1,\ln}{F}_{ab}  + \ov{1}{F}_{ab} )+o(r^{-1}) ,
\end{array}
\ee
where one can compute the various field strength coefficients in terms of the gauge field coefficients.\footnote{In the main body of the paper we use the notation $F_{ru} \equiv \ov{0}{F}_{ru},   F_{ra} \equiv \ov{0}{F}_{ra}, F_{ua}  \equiv   \ov{0}{F}_{ua}$.}
We now consider the field and gauge fixing equations. Starting with the later, we find
\begin{multline} \label{exprharmgge}
0=-\nabla^\mu \A_\mu = \partial_u \A_r+ r^{-2}\partial_r(r^2(\A_u-\A_r))-r^{-2}\partial^b \A_b   \\ =\frac{\ln r}{r^2}(\partial_u \ov{0, \ln}{A}_r + \ov{0,\ln}{A}_u)  + \frac{1}{r^2}(\partial_u \ov{0}{A}_r + \ov{0,\ln}{A}_u-\partial^b A_b)  + \frac{\ln r}{r^3}(\partial_u \ov{1,\ln}{A}_r +2 \ov{1, \ln^2}{A}_u- \partial^b \ov{1, \ln}{A}_b )  \\
+ \frac{1}{r^3}(\partial_u \ov{1}{A}_r+\ov{1, \ln}{A}_u - \ov{0, \ln}{A}_r- \partial^b \ov{1}{A}_b) + o(r^{-3}).
\end{multline}
We will think of these equations as determining the various coefficients of $\A_r$ in terms of the remaining components of the gauge field. The vanishing of the $O(\ln r/r^2)$ factor in \eqref{exprharmgge} implies\footnote{Notice that the non-abelian contribution to $\F_{ru}$ starts at order $\ln^2/r^3$ and hence it does not appear in \eqref{Fru0ln} and \eqref{Fru0}.}
\be \label{Fru0ln}
 \ov{0,\ln}{F}_{ru} = -\ov{0,\ln}{A}_u - \partial_u \ov{0,\ln}{A}_r =0.
\ee
 We next consider the field equation \eqref{fieldeq} for $\nu=u$:
\begin{multline} \label{ufieldeq}
-\D^\mu \F_{\mu u} =  \D_u \F_{ru}- r^{-2}\D_r (r^2 \F_{ru})+ r^{-2}q^{ab} \D_a \F_{ub} =
\frac{1}{r^2}(\partial_u \ov{0}{F}_{ru} + D^a \ov{0}{F}_{u a} ) \\
+ \frac{\ln^2 r}{r^3}\partial_u \ov{1,\ln^2}{F}_{ru}  + \frac{\ln r}{r^3}(\partial_u \ov{1,\ln}{F}_{ru} + [\ov{0,\ln}{A}_u, \ov{0}{F}_{ru}]+ D^a \ov{1,\ln}{F}_{ua} +  [\ov{1,\ln}{A^a} , \ov{0}{F}_{ua}]) \\
+ \frac{1}{r^3}(\partial_u \ov{1}{F}_{ru} + D^a \ov{1}{F}_{ua} +  [\ov{1}{A^a},   \ov{0}{F}_{ua}]) +o(r^{-3})=0.
\end{multline}
The resulting equations can be used to determine the coefficients of $\F_{ru}$ in terms of previously determined data. This in turn fixes the $\A_u$ coefficients. For instance, consider the expression for  $\ov{0}{F}_{ru}$ in terms of the gauge field coefficients,
\ba \label{Fru0}
\ov{0}{F}_{ru} & =&\ov{0,\ln}{A}_u - \partial_u \ov{0}{A}_r \\
& =& 2 \ov{0,\ln}{A}_u - \partial^b A_b ,
\ea
where in the second equality  we used the $O(1/r^2)$ gauge fixing condition \eqref{exprharmgge}. Using that $\ov{0}{F}_{u a} = \partial_u A_a$, the $O(1/r^2)$ equation in \eqref{ufieldeq} implies
\be
\partial_u \ov{0,\ln}{A}_u = - \frac{1}{2}[A^b, \partial_u A_b],
\ee
which determines $\ov{0,\ln}{A}_u$ in terms of the free data $A_a$ (given boundary conditions in $u$ as discussed in the main text). Going to the next order, we see that \eqref{ufieldeq} implies $\partial_u \ov{1,\ln^2}{F}_{ru}=0$. We assume the stronger condition
\be
0=\partial_u \ov{1,\ln^2}{F}_{ru} = -2 \ov{1,\ln^2}{A}_u + [ \ov{0,\ln}{A}_r,  \ov{0,\ln}{A}_u],
\ee
from which we obtain $\ov{1,\ln^2}{A}_u $ in terms of  previously determined coefficients. To solve for higher order coefficients of $\A_u$ using \eqref{ufieldeq} requires knowledge of lower order $\A_a$ coefficients. We finally discuss the remaining equations to determine them. We will consider a combination of the $\nu=a$ field equation
\be \label{afieldeq}
\D^\mu \F_{\mu a} =  -\D_u \F_{ra} + \D_r\F_{ra}-\D_r \F_{ua} + r^{-2} q^{bc} \D_c \F_{ba}  =0
\ee
with the $(r,u,a)$ Bianchi identity
\be  \label{ruabianchi}
\D_r \F_{ua} + \D_u \F_{ar}+ \D_a \F_{ru}=0.
\ee
 When adding \eqref{afieldeq} and \eqref{ruabianchi}, the $\F_{ua}$ terms cancels and one gets
\begin{multline}
0 = - 2 \D_u \F_{ra} + \D_r \F_{ra}  + \D_a \F_{ru} + r^{-2} q^{bc} \D_c \F_{ba}  =
\frac{\ln r}{r^2}(-\partial_u \ov{0,\ln}{F}_{ra})\\
+ \frac{1}{r^2}(-2 \partial_u \ov{0}{F}_{ra}   +D_a \ov{0}{F}_{ru} + D^b F_{ba}   ) +o(r^{-2}).
\end{multline}
The leading order condition implies $-\partial_u \ov{0,\ln}{F}_{ra}=0$. Again, we will assume the stronger condition\footnote{In the context of loop-level subleading soft theorem, the matching of charges at spatial infinity requires a non-trivial, $u$-independent $\ov{0,\ln}{F}_{ra}(x)$ coefficient \cite{sayali1,sayali2}.}
\be
0=\ov{0,\ln}{F}_{ra}= - \ov{1,\ln}{A}_a - D_a \ov{0,\ln}{A}_r,
\ee
which determines $\ov{1,\ln}{A}_a$. This now allows one to continue one more order in Eqs. \eqref{exprharmgge}, \eqref{ufieldeq} to determine $\ov{1,\ln}{A}_r$ and $\ov{1,\ln}{A}_u$ . The procedure can be continued to higher orders but for the purposes of this paper the displayed relations are sufficient.

\section{Residual gauge parameters}\label{ggeparameterapp}
\subsection{Asymptotic expansion} 
In this appendix we discuss the asymptotic expansion of residual $O(r^0)$ and $O(r)$ gauge parameters. To subleading order in the $r \to \infty$ expansion, one finds
\ba
\Lambda^0_\lambda(r,u,x) & =& \lambda(x) + \frac{\ln r}{r}\ov{\ln}{\lambda}(u,x) + O(\ln^2 r/r^2) ,   \label{Lam0app}\\
\Lambda^1_\e(r,u,x) & = & r \e (x) + \ln r \ov{0,\ln}{\e}(u,x) + \ov{0}{\e}(u,x)+ O(\ln^3 r/r). \label{Lam1app} 
\ea
As illustrated below, the subleading coefficients are determined by recursively solving the residual gauge equation \eqref{ggeeq}
\begin{multline} \label{ggeeqapp}
\square \Lambda + [\A_\mu ,   \nabla^\mu \Lambda]  = r^{-1}(\partial^2_r -2 \partial_u \partial_r +r^{-2} \partial^a \partial_a) (r \Lambda) \\ +[\A_r-\A_u,\partial_r \Lambda]-[\A_r , \partial_u \Lambda] + r^{-2}q^{ab}[\A_a,\partial_b \Lambda] =0,
\end{multline}
where recall that $\partial_a$ is used to denote the sphere covariant derivative and hence $\partial^a \partial_a$ is the sphere laplacian. 
Substituting \eqref{Lam0app} and \eqref{fallrAapp} in \eqref{ggeeqapp} one finds
\be \label{resggeLam0}
\square \Lambda^0_\lambda + [\A_\mu ,   \nabla^\mu \Lambda^0_\lambda] = \frac{1}{r^2}(-2 \partial_u \ov{\ln}{\lambda} + D^a \partial_a \lambda) + O(\ln^2 r /r^3).
\ee
Similarly for \eqref{Lam1app} one gets
\begin{multline} \label{resggeLam1}
\square \Lambda^1_\e + [\A_\mu ,   \nabla^\mu \Lambda^1_\e] = \frac{\ln r}{r}(-2 \partial_u \ov{0,\ln}{\e} -[\ov{0,\ln}{A}_u,\e] ) \\
+\frac{1}{r}(-2 \partial_u \ov{0}{\e}-2 \partial_u \ov{0,\ln}{\e} +(\partial^a \partial_a +2) \e+ [A^a,\partial_a \e]) + O(\ln^2 r /r^2).
\end{multline}
The vanishing of the factors with a given $r$-dependence in \eqref{resggeLam0} and \eqref{resggeLam1} can be used to determine the subleading coefficients in terms of the respective leading coefficients $\lambda$ and $\e$. 

We finally describe the residual $O(r^0)$ gauge parameter in the extended space \eqref{defGamma}, which is required   to  compute the modified bracket \eqref{modbrack} between   $O(r^0)$ and $O(r)$ gauge parameters.  The residual gauge equation in the extended space is
\be\label{ggeeqtilde}
\nabla^\nu \tilde{\D}_\mu \tilde{\Lambda} =0
\ee
where  $\tilde{\D}_\mu$ is the  $\tilde{\A}_\mu$--gauge covariant derivative, which in the parametrization \eqref{defGamma} reads
\be
 \tilde{\D}_\mu = \D_\mu + [\D_\mu \Lambda^1_\phi, \cdot ].
 \ee
Since we are working to first order in $\phi$, it suffices we consider a corrected gauge parameter of the form
\be
 \Lamt^0_\lambda = \Lambda^0_\lambda + \Lambda^0_{\lambda,\phi},  \label{Lamlamphi}
\ee
where $\Lambda^0_{\lambda,\phi}=O(\phi)$ is the correction. Substituting \eqref{Lamlamphi} in \eqref{ggeeqtilde} and  imposing the equation to to first order in $\phi$ one finds $\Lambda^0_{\lambda,\phi}$ must satisfy
\be
\square \Lambda^0_{\lambda,\phi} + [\A_\mu ,   \nabla^\mu \Lambda^0_{\lambda,\phi}]   + [\D_\mu \Lambda^1_\phi, \nabla^\mu \Lambda^0_\lambda]=0.
\ee
This equation is of the same form as the original residual gauge equation \eqref{ggeeqapp} except that it now has a ``source'' term.  Proceeding as in the earlier cases, one finds 
\be
 \Lambda^0_{\lambda,\phi} = \ln r  \ov{0,\ln}{\lambda}_{\phi} + \ov{0}{\lambda}_{\phi} +o(1)
\ee
with the coefficients determined by the equations
\be
\partial_u \ov{0,\ln}{\lambda}_{\phi} = - \frac{1}{2} [\phi, \partial_u \ov{\ln}{\lambda}] , \quad \quad \partial_u \ov{0}{\lambda}_{\phi} =   \frac{1}{2} [\phi, \partial_u \ov{\ln}{\lambda}] + \frac{1}{2} [D^a \phi, \partial_a \lambda].
\ee

\subsection{Commutators}
Having understood the $O(r^0)$ gauge parameter in the extended space, we can now compute its modified bracket \eqref{modbrack} with the $O(r)$ gauge parameter:
\ba
[\Lambda^1_\e , \Lamt^0_\lambda]^* & =& [\Lambda^1_\e,\Lamt^0_\lambda] + \delta_{\Lambda^1_\e} \Lamt^0_\lambda - \delta_{\Lamt^0_\lambda} \Lambda^1_\e,  \\
& =&  [\Lambda^1_\e,\Lambda^0_\lambda] + \Lambda^0_{\lambda, \e} -  \delta_{\Lambda^0_\lambda} \Lambda^1_\e ,  \label{modbracket12} \\
& =&  \Lambda^1_{[\e,\lambda]}, \label{10bracket}
\ea
In going from the first to second line,  we used that $\delta_{\Lambda^1_\e} \Lambda^0_\lambda=0$ and $\delta_{\Lambda^1_\e} \Lambda^0_{\lambda,\phi}=  \Lambda^0_{\lambda,\e}$, as follows from \eqref{del1e}, and dropped  $O(\phi \e)$ terms. The last equality can be understood by a similar argument as for the $O(r^0)$ bracket \eqref{Lam0bracket}: By construction,  \eqref{modbracket12} satisfies the gauge parameter equation \eqref{ggeeq}. Furthermore, the leading $r \to \infty$ behavior of \eqref{modbracket12} is captured by the  commutator $[\Lambda^1_\e,\Lambda^0_\lambda] \sim r [\e,\lambda]+ \cdots$.  These are precisely the two defining conditions for $\Lambda^1_{[\e,\lambda]}$. One can also explicitly check the equality $[\Lambda^1_\e , \Lamt^0_\lambda]^*= \Lambda^1_{[\e,\lambda]}$ in the $r \to \infty$ limit by using the expansions given in the previous subsection. By similar arguments one can verify that $[\Lamt^0_\lambda, \Lamt^0_{\lambda'}]^*=  \Lamt^0_{[\lambda,\lambda']}$.

\subsection{Bulk description of $O(r^0)$ parameters in the extended space}
In the body of the paper we defined the extension of $O(r^0)$ large gauge transformations to $\Gext$ by  \eqref{del0lambda}. In the notation of \eqref{defGamma} this translates to
\be \label{dellamAt}
 \delta_{\lambda} \At_\mu = \D_\mu \Lambda^0_\lambda +\D_\mu \Lambda^1_{[\phi,\lambda]}.
\ee
However, from the ``bulk'' description \eqref{defGamma}  it may be  more natural to instead consider
\ba
\tilde{\delta}_\lambda \At_\mu := \tilde{\D}_\mu  \Lamt^0_\lambda & = & \D_\mu \Lambda^0_\lambda +\D_\mu \Lambda^0_{\lambda,\phi}+ [\D_\mu \Lambda^1_\phi,\Lambda^0_\lambda  ]   \label{DtLamt0} \\
& = &    \D_\mu \Lambda^0_\lambda + \D_\mu \Lambda^1_{[\phi,\lambda]} + \delta_{\Lambda^0_\lambda}\D_\mu \Lambda^1_\phi, \\
& = &   \delta_{\lambda} \At_\mu  + \delta_{\Lambda^0_\lambda}\D_\mu \Lambda^1_\phi  \label{DtLamt02}
\ea
where the second equality follows from evaluating $\D_\mu$ on both sides of the equality in \eqref{10bracket}, for $\e=\phi$. We thus see that  the alternative  extension differs from \eqref{dellamAt} by the last term in \eqref{DtLamt02}.  This term however  does not respect the form of $\tilde{A}_\mu$ given in  \eqref{defGamma}. Thus, to use the extension \eqref{DtLamt0} one would have to work in a further enlarged space so as to allow such kind of terms. We expect such treatment should  lead to results consistent with the ones obtained in this paper with the simpler extension \eqref{dellamAt}.

\section{Types of corrections to asymptotic charges} \label{softthmsapp}

Throughout the paper we have made reference to the link between asymptotic charges and  soft gluon theorems. In this appendix we discuss some aspects of this relationship.

Let $\S_n$ be a  scattering amplitude involving $n$ external gluons. Its dependence with the YM coupling $g$ is of the form
\be
\S_n = g^{n-2}( \S_n^\tree + g^2 \S_n^\loop + \cdots).
\ee
On the other hand, the dependence of the asymptotic charges on $g$ can be obtained from the expressions given in the body of the paper (where we set $g=1$) by doing the replacements $A_a \to g A_a$, $\Lambda \to g \Lambda$ and multiplying by an overall $g^{-2}$ factor.  One then finds that both $Q^{0, \rad}_\lambda$ and $Q_Y$ depend on $g$ according to
\be \label{Qtree}
Q^\tree = Q^{\soft (1)} + g Q^{\hard (2)},
\ee
where $Q^{\soft (1)}$ is linear in the gauge field  and inserts a soft ($\w \to 0$ energy) gluon and $Q^{\hard (2)}$ is quadratic in the gauge field and preserves the number of external gluons. Schematically one has
\be \label{Qsoft1Qhard2}
[Q^{\soft (1)}, \S_n^\tree ] \sim g \S^\tree_{n+1_s}, \quad  [Q^{\hard (2)}, \S_n^\tree ] \sim \S_n^\tree,
\ee
where the power of $g$ in the first equation comes from the extra coupling needed to get a non-trivial amplitude with $n+1_s$ gluons. The ``$s$'' label indicates that the extra gluon is soft.  The (leading or subleading) single-soft tree-level factorization theorem can then be written as
\be \label{treeWard}
[Q^\tree , \S^\tree_n ] \sim 0. 
\ee

In our analysis we encountered two kinds of corrections the charges \eqref{Qtree} may receive. 

The first type of correction are ``hard'' terms that are of higher order in the gauge field. For instance
\be \label{Qextrahard3}
Q= Q^\tree + g^2 Q^{\hard (3)}.
\ee 
When acting on the S matrix this cubic correction would yield, to lowest order, a term of the form
\be \label{extrahard}
[ g^2 Q^{\hard (3)} , \S_n ] \sim g^n [Q^{\hard (3)} ,\S_n^\tree ] \sim g^{n+1}  \S_{n+1}^\tree , 
\ee
where the $O(A^3)$ charge adds an external gluon (not necessarily soft), with the corresponding extra $O(g)$ factor.  
Since the coupling power corresponding to \eqref{treeWard} is $g^{n-1}$, the  term \eqref{extrahard} does not affect the tree-level Ward identity.  At 1-loop however, the  Ward identity  would get contributions from \eqref{extrahard} and from 
\be \label{QtreeSloop}
[Q^\tree , g^n \S_n^\loop] \sim g^{n+1}(\S_{n+1_s}^\loop + \S_n^\loop).
\ee
Thus, in the case of a charge of the form \eqref{Qextrahard3}, the would-be Ward identity would enforce the sum of \eqref{extrahard} and \eqref{QtreeSloop} to vanish. The study of such potential identities is however well-beyond the scope of the present paper.  Even to make sense of such expressions would require a  treatment of infrared divergences that are not present in the tree-level case.

There is a second type of correction the charge may acquire, of the type
\be \label{Qextrasoft2}
Q= Q^\tree + g Q^{\soft (2)},
\ee 
which is quadratic in the soft part of the gauge field. As discussed in \cite{cl} in the gravitational context, such kind of terms are non-trivial if at least one of the external states is  soft. Schematically,
\be
[Q^{\soft (2)} , \S^\tree_{n_h} ] =0, \quad  \text{but} \quad [Q^{\soft (2)} , \S^\tree_{n_h+1_s} ] \sim  \S^\tree_{n_h+1_s} .
\ee
Thus, a would-be Ward identity with charge \eqref{Qextrasoft2} would lead to a relation of the type
\be
0=[ Q, \S^\tree_{n_h+1_s}] \sim g(\S^\tree_{n_h+2_s} + \S^\tree_{n_h+1_s}) .
\ee
Such kind of terms are thus sensitive to (tree-level) double-soft gluon emission. See \cite{cl,anupam,distler} for related discussion of double-soft emission and asymptotic charges.


\section{Covariant LPS charge} \label{covLPSapp}
We start by expanding the covariant derivatives in  \eqref{QYcov} as 
\ba \label{integrand}
\partial_u (D_a F_{ru} + D^b F_{ba}) & = & \partial_a \partial_u F_{ru} + \partial^b \partial_u F_{ab} + \partial_u( [A_a,F_{ru}]+ [A^b, F_{ba}] ).
\ea
Recalling the expressions for $F_{ru}$ \eqref{FruJu} and $F_{ab}$ \eqref{FabitoAa},  one can see that the first two terms in the rhs of \eqref{integrand} lead to $Q^\soft_Y$ plus the $J_u$ part of $Q^\hard_Y$ plus an extra term coming from the non-abelian part of $F_{ab}$,
\be \label{defQextraY}
Q^\extra_Y := \int du d^2 x  u \tr Y^a \partial_u \partial^b [A_a,A_b].
\ee
The last two terms of \eqref{integrand} lead the $J_a$ part of $Q^\hard_Y$ plus cubic terms plus quadratic ``soft'' terms given by
\ba
Q^{(2)}_Y & = & \int d^2 x  \tr Y^a [ E^1_a,A^+_a]\\
Q^{(3)}_Y & = & \int du d^2 x  u \tr Y^a \partial_u \big( [ A_a,   \int^\infty_u J_{u'} du' ] + [A^b,[A_b,A_a]] \big)
\ea
where
\be
E^1_a(x) := \int^\infty_{-\infty} du  u \partial_u A_a(u,x).
\ee
Summarizing, the charge $Q^\cov_Y$ defined by \eqref{QYcov} differs from the LPS charge $Q_Y$ by the three terms:
\be
Q^\cov_Y-Q_Y = Q^\extra_Y + Q^{(2)}_Y +Q^{(3)}_Y .
\ee
As explained in  appendix \ref{softthmsapp}, $Q^{(2)}_Y$ and $Q^{(3)}_Y$ yield trivial single-insertion tree-level Ward identities. On the other hand $Q^\extra_Y$  produces non-trivial terms in the single Ward identity that are in conflict with the subleading soft gluon theorem. To see the form of these terms, let us evaluate the commutator $Q^\extra_Y$ with $\Ah_a(\w,x)$.  For the purposes of this evaluation, one can assume decaying $u$ fall-offs in $A_a(u,x)$ and integrate by parts the $u$-derivative in \eqref{defQextraY}. It is also convenient to integrate by parts in the sphere to work with the expression 
\be
Q_Y^\extra = - \frac{1}{2} \int du d^2 x   \tr (\partial_a Y_b- \partial_b Y_a)  [A^a,A^b].
\ee
 From the elementary commutator
 \be
 [A^\beta_b(u',x'),\Ah^\alpha_a(\w,x)]_{\textrm{op}} = - \delta^{\alpha \beta}q_{ab} \frac{e^{i \w u}}{2 \w} \delta^{(2)}(x',x)
 \ee
(where $\alpha$ and $\beta$ are color indices) one finds
\be \label{extrahardterm}
[Q_Y^\extra, \Ah_a(\w,x)]_{\textrm{op}} = - \frac{1}{2 \w} [ \partial_a Y_b- \partial_b Y_a , \Ah^b(\w,x)].
\ee
Thus, a tree-level Ward identity for $Q^\cov_Y$ would include the above  contribution in addition to the terms \eqref{defdelY}. Since the latter already capture the tree-level subleading soft factors, such Ward identity would be in contradition with the soft theorem, except for ``electric'' vector fields $Y_a=\partial_a \e$ for which the extra term is absent. Whereas this suffices for the purposes of the present paper, there is more to be understood if one wishes  to include  ``magnetic'' charges with $Y_a= \epsilon_{a}^{\ b}\partial_b \mu$. See  footnote \ref{magfoot} for further comments on this issue.

\section{Comparison with the gravitational case} \label{gravsec}
In the gravitational case, the standard radiative fall-off conditions on the spacetime metric take the form (focusing for simplicity on the angular components of the metric)
\be \label{gab}
g_{ab} \stackrel{r \to \infty}{=} r^2 q_{ab} + r C_{ab} + \cdots 
\ee
where $q_{ab}$ is the $u$-independent 2d metric on the celestial sphere and $C_{ab}=C_{ab}(u,x)$ encodes  gravitational radiation at null infinity. The latter plays the role of ``free data'' that determines the asymptotic metric components through Einstein equations (supplemented by gauge fixing conditions).  Schematically, 
\be
\Ggravrad \approx \{ C_{ab}(u,x)  \}.
\ee
The non-trivial diffeomorphisms preserving \eqref{gab} are asymptotic Lorentz transformations plus so-called supertranslations, generated by vector fields of the form
\be
\xi^0_f = f(x) \partial_u + \cdots,
\ee
with arbitrary $f(x)$. These can be thought of as the analogous to the $O(r^0)$ large gauge transformations in the YM case. The analogy however breaks down at the level of the algebra, since supertranslations are abelian,
\be
[\xi^0_f, \xi^0_{f'}]^* =0
\ee
(the bracket is now  a modified vector field Lie bracket analogous to \eqref{modbrack}, see e.g. \cite{Barnich:2013sxa}).

The above asymptotic symmetries can be enlarged to include  superrotations, generated by vector fields with a leading non-trivial angular component
\be \label{xiY}
\xi^1_Y = Y^a(x) \partial_a + \cdots
\ee
with arbitrary $Y^a(x)$.\footnote{In the special case where $Y^a$ is a (global) conformal Killing vector field of $q_{ab}$, these are not new symmetries but represent the generators of the asymptotic Lorentz group. Although we shall be phrasing the extension in terms of arbitrary sphere vector fields, the following discussion applies equally well to the case of local conformal Killing vector fields.} 
 However, to allow for such transformations one needs to relax the form of the asymptotic metric \eqref{gab}. 
As in the YM case, one can write a linearized version of the extended space as:
\be \label{Ggravext}
\Ggravext = \{ \tilde{g}_{\mu \nu} = g_{\mu \nu} + \L_{\xi^1_X} g_{\mu \nu} , \quad g_{\mu \nu} \in \Ggravrad,  \quad X^a \in \mathfrak{X}(S^2) \} . 
\ee
Of course, in the gravitational case we known how to ``exponentiate'' the linearized extension in order to obtain a full extended space. We will however continue the discussion within the linearized setting in order to make contact with the analysis in this paper. 

The space \eqref{Ggravext} is then naturally parametrized by pairs
\be \label{Ggravext1}
 \Ggravext \approx \{ (C_{ab}(u,x), X^a(x))  \},
\ee
where $C_{ab}$ ``generates'' the metric $g_{\mu \nu}$ and $X^a$ implements its superrotated version.\footnote{Strictly speaking, $X^a$ should be considered modulo global conformal Killing vector fields on the celestial sphere. Other more commonly used labels for the ``superrotation frame'' are either the deformed 2d metric $\tilde{q}_{ab}$ or the so-called Geroch/Liouville tensor $T_{ab} =-(D_a D_b)^\tf D_c X^c$,  see e.g. \cite{Barnich:2010eb,clnewsym,comperevacua}.

}  Eqs. \eqref{Ggravext}, \eqref{Ggravext1} are the gravitational versions of Eqs. \eqref{defGamma}, \eqref{Gammavariables} in the YM case.  The extended gravitational space now supports the action of superrotations \eqref{xiY}, which in the parametrization \eqref{Ggravext1} takes the form
\be \label{del1Y}
\delta^1_Y C_{ab} = 0 , \quad  \delta^1_Y X^a   = Y^a,
\ee
in parallel to Eq. \eqref{del1e}.  The trivial action on $C_{ab}$ in \eqref{del1Y} may appear at odds with the standard action of superrotations one encounters in the literature. The reason is that the definition of $C_{ab}$ in \eqref{Ggravext1} does not agree with the usual one. In order to compare with the more standard  parametrization, let us look at the angular components of the full metric $\tilde{g}_{\mu \nu}$ in \eqref{Ggravext}. These take the form
\be
\tilde{g}_{ab} = r^2 \tilde{q}_{ab}+ r \tilde{C}_{ab} + \cdots,
\ee
where
\ba
\tilde{q}_{ab} & =& q_{ab} + X \cdot q_{ab},  \\
\tilde{C}_{ab} & =& C_{ab}+ X \cdot C_{ab} - u (D_a D_b)^\tf D_c X^c , \label{tildeC}
\ea
($\tf$ stands for trace-free part)  with
\ba
X \cdot q_{ab} & :=& (\L_X - D_c X^c) q_{ab} \\
X \cdot C_{ab} & :=& (\L_X + \frac{1}{2}D_c X^c( u \partial_u -1) ) C_{ab} .
\ea
In the standard parametrization one treats $\tilde{C}_{ab}$, rather than $C_{ab}$, as independent variable.\footnote{What we are calling  $\tilde{C}_{ab}$ is  what  usually is  denoted as $C_{ab}$ (e.g.  \cite{Barnich:2010eb,stromvirasoro,superboost}).} From \eqref{tildeC} and \eqref{del1Y} one finds
\be \label{del1YtildeC}
\delta^1_Y \tilde{C}_{ab} =  Y \cdot \tilde{C}_{ab} - u (D_{a} D_{b})^\tf D_c Y^c ,
\ee
which coincides with the standard superrotation action (recall we are working to linear order in $X^a$ and $Y^a$ and dropping  $O(X^2)$,  $O(Y^2) $ and $O(X Y)$ terms). In this way, we can think of $C_{ab}$ as a ``dressed'' version of the standard $\tilde{C}_{ab}$ which neutralizes the action of superrotations. 

There is yet another parametrization that is useful to consider, which consists in removing  the $O(u)$ part of $\tilde{C}_{ab}$, 
\ba
\hat{C}_{ab} &: =& \tilde{C}_{ab}  - u (D_a D_b)^\tf D_c X^c  \\
 &=& C_{ab}+ X \cdot C_{ab}. \label{ChatitoC}
\ea
In this case the action of superrotations is purely homogeneous,
\be \label{del1YtildeC}
\delta^1_Y \hat{C}_{ab} =  Y \cdot \hat{C}_{ab}.
\ee

Regardless of the parametrization being used, the algebra relation between supertranslations and superrotations is given by 
\be \label{xioxi1}
[\xi^0_f , \xi^1_Y]^* = - \xi^0_{Y \cdot f}, 
\ee
where
\be
Y \cdot f := (\L_Y - \frac{1}{2}D_c Y^c) f.
\ee
The algebraic relation \eqref{xioxi1} is again structurally different from its YM counterpart, in that the RHS is  an $O(r^0)$ rather than $O(r)$ symmetry. 
Schematically, 
\be \label{xi01Lam01}
[\xi^0 , \xi^1] \sim \xi^0 \quad \text{vs.} \quad [\Lambda^0,\Lambda^1] \sim \Lambda^1 .
\ee

Let us now discuss charges. In the  radiative space, supertranslations are generated by 
\be
P^\rad_f = \int d^2 x f(x) \P(x)
\ee
where $\P$ is the supermomentum density defined by
\be \label{defcalP}
 \P = \int^\infty_{-\infty} du (N^{ab} N_{ab} -2 D^a D^b N_{ab}),
\ee
and $N_{ab} \equiv \partial_u C_{ab}$. Following the same logic as for the YM case, the equation that determines the extension of this charge from $\Ggravrad$ to $\Ggravext$ is 
\be
\delta^1_Y P^\ext_f  = - P^\ext_{Y \cdot f}.
\ee
This has the simple solution
\ba \label{Pf1}
P^\ext_f & = & P^\rad_f - P^\rad_{X \cdot f}  , \\
&=& \int d^2 x f ( \P + X \cdot \P) \label{Pf2},
\ea
where to get to the second line we integrated by part on the sphere, and
\be
X \cdot \P = (\L_X  + \frac{3}{2}D_c X^c ) \P.
\ee
Eq. \eqref{Pf1} is the analogous to Eq. \eqref{defQ0lambda}. The main difference with the YM case is that the extension is determined from the $\xi^0$ charge itself, rather than from the $\xi^1$ one. This is a consequence of the difference displayed in \eqref{xi01Lam01}.

In order to compare \eqref{Pf1} with the known expression of supertranslation charges, let us rewrite \eqref{Pf1} in terms of the tensor $\hat{N}_{ab}\equiv \partial_u \hat{C}_{ab}$. In the $O(X)$ setting we are working, Eq. \eqref{ChatitoC} implies
\be \label{NitoNhat}
N_{ab}= \hat{N}_{ab} - X \cdot \hat{N}_{ab}.
\ee
Substituting \eqref{NitoNhat} in \eqref{defcalP}, \eqref{Pf2} and neglecting $O(X^2)$ terms one finds
\be \label{calPext}
\P + X \cdot \P = \hat{\P} - \hat{N}^0_{ab} D^a D^b D_c X^c ,
\ee
where $\hat{\P}$ is given by \eqref{defcalP} with $N_{ab}$ replaced by $\hat{N}_{ab}$ and
\be
\hat{N}^0_{ab} := \int^\infty_{-\infty} du \hat{N}_{ab}.
\ee
The RHS of Eq. \eqref{calPext} can be seen to  coincide with the standard (extended) supermomentum density. In particular, the last term in \eqref{calPext} is the known modification  that features the Geroch/Liouville tensor $T_{ab} =-(D_a D_b)^\tf D_c X^c$, see \cite{cp} and references therein.

One can similarly construct superrotation charges. The situation is again  structurally different from the YM case due to \eqref{xi01Lam01}. In particular, the difficulties we encountered in  YM to construct a covariant $Q^1_\e$ have no analogue for superrotations. Nevertheless, the algebraic requirement coming from \eqref{xioxi1} imposes subtle conditions on the form of the superrotation charge, see \cite{cp} for details.

To conclude, the differences between the YM and gravity cases may be summarized by saying that the notion of 
 ``covariance'' at null infinity in each theory is given by transformations associated to different radial/energy order: In  YM  the  natural notion of covariance comes from  $\Lambda^0$ transformations, whereas in gravity it comes from $\xi^1$ transformations. 




\end{document}